\documentclass[12pt]{article}
\usepackage{amsmath,amssymb,amsopn,amsthm,times}

\title{Feynman problem in the noncommutative case}

\author{
Jos\'e F. Cari\~nena\dag\ and
H\'ector Figueroa\ddag
\\[1pc]
\dag\,Departamento de F\'{\i}sica Te\'orica, 
Universidad de Zaragoza,\\
50009 Zaragoza, Spain
\\[1pc]
\ddag\,Departamento de Matem\'aticas, Universidad de Costa Rica,\\
2060 San Pedro, Costa Rica}

\scrollmode
 
\advance\topmargin by -\headheight
\advance\topmargin by -\headsep
\evensidemargin=\oddsidemargin
\theoremstyle{plain}

\theoremstyle{definition}

\DeclareMathOperator{\Div}{div}    
\newcommand{\dl}{\delta}           
\newcommand{\eps}{\varepsilon}     
\newcommand{\F}{\mathcal{F}}       
\def\fd#1#2{\frac{d#1}{d#2}}       
\newcommand{\Ga}{\Gamma}           
\newcommand{\La}{\Lambda}          
\def\pd#1#2{\frac{\partial#1}{\partial#2}}
\newcommand{\w}{\wedge}            
\newcommand{\x}{\times}            
\renewcommand{\:}{\colon}          
\def\<#1,#2>{\langle#1,#2\rangle}  

\makeatletter
\def\section{\@startsection{section}{1}{\z@}{-3.5ex plus -1ex minus
	 -.2ex}{2.3ex plus .2ex}{\large\bf}}
\def\subsection{\@startsection{subsection}{2}{\z@}{-3.25ex plus -1ex
	 minus -.2ex}{1.5ex plus .2ex}{\normalsize\bf}}
\makeatother

\InputIfFileExists{Whatnot.tex}{}{} 

\begin{document}

\maketitle

\begin{abstract}
In the context of the Feynman's derivation
of electrodynamics, we show that 
noncommutativity allows other particle
dynamics than the standard formalism of 
electrodynamics.
\end{abstract}

\medskip

\noindent \textit{Keywords}: Feynmann problem, minimal coupling, noncommutative spaces.

\noindent PACS numbers: 11.10.Nx, 45.20.Jj.

\section{Introduction}

Feynman procedure~\cite{DysonFey} to obtain Maxwell's 
equations in electrodynamics has been reviewed under
different kind of settings, and several nontrivial and
interesting generalizations are possible, see for  
instance~\cite{CIMS,Lee,Tanimura,Bracken,Silagadze,Berard,Paschke}.  
In general the locality property that different coordinates commute
is assumed. However, as pointed out by Jackiw~\cite{ja02}, 
Heisenberg suggested in a letter to Peierls~\cite{heisen} that 
spatial coordinates may not commute, Peierls communicated the 
same idea to Pauli \cite{pauli}, who told it to Oppenheimer;
eventually the idea arrived to Snyder \cite{snyder}
who wrote the first paper on the subject. On the other hand, 
the existence of a minimal length beyond which no strict 
localization is possible, the importance of the physics 
in noncommutative planes, the noncommutative Landau problem, 
Peierls substitution, and the fact that noncommutative field 
theory is relevant not only in string theory but also in 
condensed matters, motivated a new interest on the subject 
during the last years.
 
Due to this increasing interest in 
noncommutative field theories, it is worthwhile to consider
the noncommutative version of such procedure, where
locality no longer holds, which has a better chance 
to find new kinds of particle dynamics, which after all, 
according to Dyson~\cite{DysonFey}, was the original 
aim of Feynman. Such considerations were actually done 
in~\cite{moros}, but the argument given there seems to be 
 inadequate or incomplete for 
two reasons: they only considered the case where the
nonlocality is described by a coordinate independent
Moyal Bracket, whereas nowadays the non--constant
(i.e. coordinate dependent) noncommutative spaces
are gaining a lot of attention in the noncommutative 
realm, because of the appearance of such type of 
noncommutativity in various contexts specially in string
theory. Among the papers that invoke variable
noncommutativity are~\cite{Selene,DolanN,BehrS,
HashimotoT,Melpomene,AschieriBDMSW,AldrovandiSS,CalmetK}. 
On the other hand, the treatment in~\cite{moros} 
is somewhat sloppy and the main conclusions are 
not correct, as we shall point out later on.

To avoid unnecessary complications due to operator
ordering, we shall only discuss the classical analogue
of the Feynman procedure in the noncommutative case.
Accordingly, the appropriate setting would be in terms
of Poisson brackets, which is regarded as the classical
limit of the commutator of quantum observables. But we 
shall explore the different possibilities arising
from the dependence of the fundamental brackets on
the different sets of variables involved.

Let $\F(M)$ be the algebra of functions (the algebra of 
classical observables) on a manifold $M$ (the classical 
state space). A \textit{Poisson structure} on
 $M$ is a 
real skew symmetric bilinear map 
$\{\cdot,\cdot\}\:\F(M)\x\F(M)\to\F(M)$
satisfying the Jacobi identity:
$$
\{F,\{G,H\}\}+\{H,\{F,G\}\}+\{G,\{H,F\}\}=0,\quad
\forall F,G,H\in \F(M)\,,
$$ 
and such that the map $X_F =\{\cdot,F\}$ is a
derivation of the Lie algebra $\F(M)$, for each 
$F\in \F(M)$, in other words, $X_F$ is a vector field, 
usually called a \textit{Hamiltonian vector field}, and 
$F$ is said to be the \textit{Hamiltonian} of $X_F$. 
This second property, called the Leibnitz' rule,
is important as there are many examples of Lie 
algebra structures on $\F(M)$ that do not satisfy
the Leibnitz' rule. 

In particular, if $\xi^a$ denotes a set of local 
coordinates on $M$, then, using the summation index
convention,
\begin{equation}
X_F =X_F(\xi^a)\pd{}{\xi^a}=\{\xi^a,F\}\pd{}{\xi^a},\label{XHam}
\end{equation}
hence $$\{F,G\}=X_G(F)=\{\xi^a,G\}\pd{F}{\xi^a}\,.$$ Thus,
\begin{equation}
\{\xi^a,G\}=-\{G,\xi^a\}=-\{\xi^b,\xi^a\}\pd{G}{\xi^b}
=\{\xi^a,\xi^b\}\pd{G}{\xi^b},
\label{eq:localformula}
\end{equation}
and the local coordinate expression of the Poisson Bracket 
becomes
\begin{equation}
\{F,G\}=\{\xi^a,\xi^b\}\pd{G}{\xi^b}\pd{F}{\xi^a}.
\label{Poissonbracket}
\end{equation}
Therefore to compute the Poisson bracket of any pair of
functions is enough to know the fundamental Poisson 
brackets
$$
\Lambda^{ab} = \{\xi^a,\xi^b\}.
$$
Moreover, the value of $\{F,G\}$ at a point $m\in M$
does not depend on $F$ and $G$ but on $dF$ and $dG$, as explicitly 
shown in~(\ref{Poissonbracket}), hence from the
Poisson structure we get a twice contravariant skew
symmetric tensor 
$$
\Lambda(dF,dG) := \{F,G\}.
$$
Indeed, if $\bar\xi=\phi(\xi)$ is another set of
local coordinates on $M$, then,
$$
\bar \Lambda^{ab}= \{\bar\xi^a,\bar\xi^b\}
= \{\phi^a,\phi^b\}
= \{\xi^c,\xi^d\}\pd{\phi^a}{\xi^c}\pd{\phi^b}{\xi^d}
= \Lambda^{cd}\pd{\phi^a}{\xi^c}\pd{\phi^b}{\xi^d}\,,
$$
so the components of $\Lambda$ change like the local
coordinates of the twice contravariant skew symmetric
tensor with coordinate expression
$$
\Lambda= \Lambda^{ab}\pd{}{\xi^a}\w\pd{}{\xi^b}.
$$
The tensor $\Lambda$ is called a Poisson tensor.
We are using the convention that in the local expression of the
wedge product only summands whose subindex on the left hand side term 
is smaller than the subindex on the right hand side term appear.

For any function $H\in \F(M)$ the integral curves
of the  dynamical vector field 
$X_H$   are precisely determined by the solutions of the 
system of differential equations
$$
\fd{\xi^a}{t} =\{{\xi^a},H\}\,,
$$
and the dynamical evolution of a function $F$ in $M$ is given by
$$
\fd{F}{t} = \{F,H\},
$$
or in local coordinates
$$
\fd F{t} = \Lambda^{ab}\pd F{\xi^a}\pd{H}{\xi^b}.
$$
In terms of $\Lambda$ the Hamiltonian vector field 
associated to $F$ is given by
$$
X_F G= -\Lambda(dF,dG)\,, \qquad \forall G\in C^\infty (M)\,.
$$
Furthermore, the Jacobi identity is equivalent to the
vanishing of the Schouten Bracket of $\Lambda$ with
itself~\cite{CIMS}.

\section{The velocity independent case}

In this section we study the Feynman argument
in the framework of a tangent bundle, in the case 
where the bracket is nonlocal; in other words, we do not
suppose that the variables on the configuration space commute.
So we assume that the Poisson manifold $M$ is the tangent
bundle $TQ$ of a $n$-dimensional configuration space $Q$, with
local coordinates  $x^i$, $\dot x^i$, for $i=1,\ldots, n=\dim Q$. Thus
 a general Poisson bracket on $TQ$ is locally given by
$$
\{F,G\}=\{x^i,x^j\}\pd{G}{x^j}\pd{F}{x^i}
+\{x^i,\dot x^j\}\pd{G}{\dot x^j}\pd{F}{x^i}
+\{\dot x^i,x^j\}\pd{G}{x^j}\pd{F}{\dot x^i}
+\{\dot x^i,\dot x^j\}\pd{G}{\dot x^j}\pd{F}{\dot x^i}\,.
$$
Although we shall concentrate on autonomous systems, our
arguments can be extended
to more general contexts. We      first
consider a bracket such that
\begin{equation}
\{x^i,x^j\} = g_{ij}(x),
\label{eq:Pbracketx}
\end{equation}
where $g_{ij}$ is an arbitrary skewsymmetric matrix of functions, fulfilling
the constraints that a Poisson bracket satisfying the Leibniz
rule impose. In other words, we examine the possibility of a
 bracket without the locality property; a condition needed, for instance,
in a classical description of a massless 
particle~\cite{BMSSZ}. We also require
\begin{equation}
m\{x^i,\dot x^j\} = \dl_{ij},
\label{eq:Pbracketv}
\end{equation}
so this part of the Poisson bracket is the same as in the 
commutative case, considered by Feynman.

Now, the Jacobi identity
$$
\{x^i,\{x^j,\dot x^k\}\}+\{\dot x^k,\{x^i,x^j\}\}
+\{x^j,\{\dot x^k,x^i\}\}=0
$$
entails, upon using~\eqref{eq:Pbracketv}, and 
$\partial{g_{ij}}/{\partial\dot x^k}=0$, 
\begin{equation}
0=\{\dot x^k,g_{ij}\}=\{\dot x^k,x^l\}\pd{g_{ij}}{x^l}
+ \{\dot x^k,\dot x^l\}\pd{g_{ij}}{\dot x^l}
=-\frac{1}{m}\pd{g_{ij}}{x^k}.
\label{eq:gwrtx}
\end{equation}
Thus, the matrix $g_{ij}$ is a constant 
skewsymmetric $3\times 3$ matrix, and nonconstant matrices
will only be possible if one assume dependence of $g$ 
on the dotted variables, but we explore this possibility 
in the next section.
   
In other words, we are assuming that the Poisson tensor
$\La$ is given by
\begin{equation}
\La= g_{ij}\, \pd{}{x^i}\w \pd{}{x^j} 
+\frac{1}{m}\,\pd{}{x^i}\w \pd{}{\dot x^i}
+ A_{ij}(x,\dot x)\,\pd{}{\dot x^i}\w \pd{}{\dot x^j}, 
\label{eq:Ptensor}
\end{equation}
where the functions $A_{ij}$ are skewsymmetric functions 
to be determined.

To continue with Feynman's argument we further assume 
Newton's equations:
\begin{equation}
m\ddot x^j = F^j(x,\dot x),
\label{eq:Newtonbis}
\end{equation}
whose solutions  are the integral curves of the vector 
field $\Gamma$ with coordinate expression  
$$
\Gamma
=\dot x^i\,\pd{}{x^i}+\frac 1m F^j(x,\dot x)\,\pd{}{\dot x^j}\,.
$$
In other words, we assume that the equations of motion
can be written as 
\begin{align}
\fd{x^i}{t} &= \{x^i,H\} = \dot x^i,  
\label{eq:fmotioneq}\\  
\fd{\dot x^i}{t} &= \{\dot x^i,H\}
= \frac{1}{m} F^i(x,\dot x)\,,
\label{eq:smotioneq}  
\end{align}
with $\{\cdot,\cdot\}$ a Poisson bracket to be determined. 
Note however that as we assumed the nonlocality property of 
the Poisson bivector, such bivector cannot be associated to a
symplectic structure defined by a regular Lagrangian,
because the locality assumption is  equivalent to the 
vanishing of the symplectic form $\omega_L$ on a pair of 
vertical fields, which is a necessary condition for the 
existence of a regular Lagrangian \cite{Cr81,Cr83}.

Now, if we restrict ourselves to the case $Q={\Bbb R}^3$, 
we can define  a field $B$, that, in analogy with the
commutative case, we may call the magnetic field, by means of 
\begin{equation}
A_{ij}(x,\dot x)= \{\dot x^i,\dot x^j\} 
=  \frac{1}{m^2} \eps_{ijk}B_k(x,\dot x)\,.
\label{eq:defB}
\end{equation}

We require $\Gamma$ to be Hamiltonian, in particular $\Ga$ 
is a derivation of the Poisson algebra structure.
Applying  $\Gamma$ to~\eqref{eq:Pbracketv} we obtain
$$
0= m\{\dot x^i,\dot x^j\} + \{x^i,F^j\},
$$
where we use the second order condition: $\Gamma x^i=\dot x^i$, 
therefore
$$
\{x^i,F^j\} = -m\{\dot x^i,\dot x^j\} = m\{\dot x^j,\dot x^i\}
= -\{x^j,F^i\},
$$
i.e. $\{x^i,F^j\}$ is skewsymmetric and there exists 
$B_k(x,\dot x)$ such that 

\begin{equation}
\{x^i,F^j\} = - \frac{1}{m} \eps^{ijk}B_k,
\label{eq:defBbis}
\end{equation}
where $\eps^{ijk}$ denotes the fully skewsymmetric Levi--Civita tensor, for
which $\eps^{123}=1$;
so, for instance,
\begin{equation}
B_3 = -m\{x^1,F^2\} = m^2\{\dot x^1,\dot x^2\}.
\label{eq:defBbbis}
\end{equation}
Now, the Jacobi identities with one position and
two velocities entail
$$
\{x^i,B_j\} = 0,
$$
and the local expression~\eqref{eq:localformula} gives
\begin{equation}
0= \{x^i,B_j\} = g_{ik}\pd{B_j}{x^k} 
+\frac{1}{m}\pd{B_j}{\dot x^i}.
\label{eq:PbofBbis}
\end{equation}
In the commutative case, i.e.~when 
$g_{ik}\equiv 0$, \eqref{eq:PbofBbis} implies that $B_j$ is 
independent of the $\dot x$'s, but in our setting this
is not necessarily true. However, notice that, for instance
$$
\{\dot x^3,B_3\} = m^2\{\dot x^3,\{\dot x^1,\dot x^2\}\}.
$$
Thus, the Jacobi identity with three different velocities gives 
\begin{equation}
\{\dot x^i,B_i\} =0.
\label{eq:fMeq}
\end{equation}
Once again the local expression of the Poisson bracket gives
$$
m\{\dot x^i,B_j\} 
= -\pd{B_j}{x^i} + m\{\dot x^i,\dot x^k\}\pd{B_j}{\dot x^k}
= -\pd{B_j}{x^i} + \frac{1}{m}\eps_{ilk}B_k\pd{B_j}{\dot x^l},
$$
and then we can rewrite~\eqref{eq:fMeq} as  
\begin{equation}
\Div {\bf B} = -\frac{1}{m}\,{\bf B}\cdot \dot\nabla \x {\bf B},
\label{eq:fMeqbis}
\end{equation}
upon using the notation 
$\dot\nabla=(\partial/\partial{\dot x^1},\partial/\partial{\dot x^2},\partial/\partial{\dot x^3})$.
This is the equation that replaces the  Maxwell
equation $\Div {\bf B}=0$ describing the absence of monopoles in the noncommutative case.

In the particular case when the field ${\bf B}$ is independent 
of the $\dot x$'s, the previous equation~\eqref{eq:fMeqbis} 
reduces indeed to the such Maxwell equation
$$
\Div {\bf B}=0.
$$

Now, we mentioned already that ${\bf B}$ may very well depend
on the variables $\dot x$, but even if we assume that 
the field ${\bf B}$ is independent of the $\dot x$'s, 
from~\eqref{eq:PbofBbis} we see that ${\bf B}$ can still
depend on the variables $x$, since the matrix $g_{ij}$,
being a constant skewsymmetric $3\times 3$ matrix, is singular. 
Therefore, the conclusion in~\cite{moros} that the 
conditions~\eqref{eq:Pbracketx}, \eqref{eq:Pbracketv} 
and ~\eqref{eq:Newtonbis} entail static Maxwell equations 
is wrong. One of the problems in~\cite{moros} is that in 
the noncommutative space that they are using, which is 
neither explicitly defined nor described, it is not 
clear at all the meaning of the variables $\dot x$.

On the other hand, in the quest of an equation similar to 
the second Maxwell equation, we define another field ${\bf E}$, the
electric field, by $E^j = F^j -\eps_{jkl}\dot x^kB_l$.
This makes sense in the commutative case because, there, 
${\bf B}$ is certainly independent of the $\dot x$'s and, as we
shall see in a moment, \eqref{eq:defBbis} implies that ${\bf F}$ 
is at most linear in the $\dot x$'s variables,
but again this is not necessarily what happens in our
setting, even if we assume independence of ${\bf B}$ on the 
$\dot x$'s variables. Indeed, from~\eqref{eq:defBbis}
and~\eqref{eq:PbofBbis} we obtain  
\begin{align*}
\{x^i,E_j\} 
&= \{x^i,F^j -\eps_{jkl} \dot x^kB_l\} \\
&= \{x^i,F^j\} -\eps_{jkl} \{x^i,\dot x^k\}B_l 
-\eps_{jkl} \dot x^k\{x^i,B_l\}  \\
&= \{x^i,F^j\} - \frac{1}{m}\eps_{jkl}B_l=0;
\end{align*}
therefore, as claimed, in the commutative case the field  ${\bf E}$ so
defined is independent of the velocities.

Following the commutative case, we apply the vector field 
$\Gamma$ to~\eqref{eq:defBbbis} (which boils down to take the 
derivative with respect to $t$ of that equation):
\begin{align}
\dot x^l\pd{B_k}{x^l} 
+ \frac 1m F^l\pd{B_k}{\dot x^l}
&= \frac{m^2}{2}\,\eps_{ijk} \,( \{\dot x^i,F^j\}
+ \{F^i,\dot x^j\})
= m\, \eps_{ijk} \, \{F^i,\dot x^j\}  \notag\\
&= m\, \eps_{ijk} \, (\{E^i,\dot x^j\} 
+  \eps_{iln}\,\{\dot x^l,\dot x^j\}B_n
+ \eps_{iln}\, \dot x^l\{B_n,\dot x^j\}).
\label{eq:sMeq}
\end{align}
Now, the local expressions of the brackets give
\begin{align*}
m\, \eps_{ijk} \, \{E^i,\dot x^j\}
&= m\, \eps_{ijk} \, \Bigl(\{x^l,\dot x^j\} \pd{E^i}{x^l} +
\{\dot x^l,\dot x^j\} \pd{E^i}{\dot x^l}\Bigr)  \\
&= \eps_{ijk} \, \Bigl(\pd{E^i}{x^j} 
+ \frac{1}{m}\eps_{ljn} B_n \pd{E^i}{\dot x^l}\Bigr)  \\
&= \eps_{ijk} \, \pd{E^i}{x^j} 
+ \frac{1}{m} (\dl_{il}\dl_{kn} - \dl_{in}\dl_{kl})
B_n \pd{E^i}{\dot x^l}   \\
&= \eps_{ijk} \, \pd{E^i}{x^j} 
+ \frac{1}{m} \Bigl(B_k \pd{E^l}{\dot x^l}
- B_n \pd{E^n}{\dot x^k} \Bigr).
\end{align*}
Moreover,
\begin{align*}
m\, \eps_{ijk} \, \eps_{iln}\,\{\dot x^l,\dot x^j\}B_n 
&= m \, (\dl_{jl}\dl_{kn} - \dl_{jn}\dl_{kl}) \,
\{\dot x^l,\dot x^j\}B_n   \\
&=-m \, \{\dot x^k,\dot x^j\} B_j \\
&= - \frac{1}{m}\,\eps_{kjl} \, B_l B_j = 0,
\end{align*}
on account of~\eqref{eq:defB}. Also, using~\eqref{eq:fMeq},
we have 
\begin{align*}
m\, \eps_{ijk} \, \eps_{iln}\, \dot x^n \{B_l,\dot x^j\}
&= m \, (\dl_{jn}\dl_{kl} - \dl_{jl}\dl_{kn}) \,
\dot x^n \{B_l,\dot x^j\} \\
&= m\, (\dot x^n\{B_k,\dot x^n\} - \dot x^k\{B_l,\dot x^l\})
\\ 
&= m\, \dot x^n\{B_k,\dot x^n\}  \\
&=m\, \Bigl(\dot x^n \{ x^l,\dot x^n\}\pd{B_k}{\dot x^l} +
\dot x^n \{\dot x^l,\dot x^n\}\pd{B_k}{\dot x^l} \Bigr)  \\
&= \dot x^l \pd{B_k}{x^l} 
+ \frac{1}{m}\, \dot x^n \, \eps_{lnr} B_r \pd{B_k}{\dot x^l}  \\
&= \dot x^l \pd{B_k}{x^l} 
+ \frac{1}{m}\, F^l \pd{B_k}{\dot x^l}
- \frac{1}{m}\, E^l \pd{B_k}{\dot x^l}.
\end{align*}
Collecting all together, we see that \eqref{eq:sMeq} reduces to
\begin{equation}
\eps_{ijk}\pd{E^i}{x^j} = 
\frac{1}{m} \, \Bigl(E^l\pd{B_k}{\dot x^l} 
+ B_l\pd{E^l}{\dot x^k} - B_k\pd{E^l}{\dot x^l} \Bigr),
\label{eq:sMeqbis}
\end{equation} 
or in other form,
$$({\rm rot}\, {\bf E})_k+\frac 1m\left(({\bf E}\cdot{\dot \nabla})B_k
+{\bf B}\cdot \pd{{\bf E}}{\dot x^k}-\Div{\bf E}\ B_k\right)=0\,,$$
which is what replaces the Maxwell equation corresponding to Faraday's law, 
in the setting suggested at the beginning of this section.

Finally, we point out that had we assumed that the fields
${\bf B}$ and ${\bf E}$ do not depend on the $\dot x$'s (so ${\bf F}$ is actually 
 a Lorentz force),  then~\eqref{eq:fMeqbis} 
and~\eqref{eq:sMeqbis} would  exactly be  the usual Maxwell equations, 
so in the limit we have a smooth transition into the
commutative case, contrary to what is claimed in~\cite{moros}.
However, here the Lorentz force condition would be an extra 
assumption, not a consequence as in the commutative case
considered in~\cite{DysonFey}.

\section{Velocity dependent Poisson brackets}

We now return to the case where the matrix 
$g_{ij}=g_{ij}(x, \dot x)$  in~\eqref{eq:Pbracketx} also 
depends on the variables $\dot x$. Then $g_{ij}$ no longer 
need to be a constant matrix, as now~\eqref{eq:gwrtx} rather imposes  
on $g_{ij}$ the condition
$$
0=-\frac{1}{m}\pd{g_{ij}}{x^k}
+ \{\dot x^k,\dot x^l\}\pd{g_{ij}}{\dot x^l}.
$$
Moreover, the Jacobi identity on $x^i$, $x^j$ and $x^k$
reduces to
\begin{align*}
0 &=\{x^i,g_{jk}\}+\{x^k,g_{ij}\}+\{x^j,g_{ki}\}  \\
&=g_{il}\pd{g_{jk}}{x^l} + g_{kl}\pd{g_{ij}}{x^l}
+g_{jl}\pd{g_{ki}}{x^l} +\frac{1}{m}\Bigl(
\pd{g_{jk}}{\dot x^i} +\pd{g_{ij}}{\dot x^k}
+\pd{g_{ki}}{\dot x^j}\Bigr),
\end{align*}
which gives exactly one more constraint on the $g_{ij}$'s,
since the skew symmetry property of $g_{ij}$ entails that
a permutation of the indexes gives the same equation
as for $i=1$, $j=2$ and $k=3$ when the permutation is
even, and negative the expression if the permutation is odd. 
Note, however, that 
$$
\Ga g_{ij}=\Ga \{x^i,x^j\}=\{\Ga x^i,x^j\}+\{x^i,\Ga x^j\}=
\{x^i,\dot x^j\} +\{\dot x^i,x^j\}=0,
$$ 
implying that the $g_{ij}$'s are constants of the motion.

Furthermore, in the previous section we did not use
the fact that the $g$'s were constant, therefore
by the same token we obtain also for $g_{ij}(x, \dot x)$
the generalized Maxwell equations~\eqref{eq:fMeqbis}
and~\eqref{eq:sMeqbis}.
 
On the other hand, even though condition~\eqref{eq:Pbracketv} 
simplified matters quite a bit, it may be useful, in some 
settings, to modify also this condition. Thus we now address
the problem when
\begin{equation}
\{x^i, x^j\} = g_{ij}(x,\dot x),
\label{eq:GPbracketx}
\end{equation}
and
\begin{equation}
m\{x^i,\dot x^j\} = \dl_{ij} + f_{ij}(x,\dot x),
\label{eq:GPbracket}
\end{equation}
where $f_{ij}$ is another matrix
compatible with the Poisson bracket properties, which
now impose several relations among the $g_{ij}$'s and 
the $f_{ij}$'s, again the $g_{ij}$'s need not be constants.

In principle, there is no
need to impose a special condition on $f_{ij}$, but
the parallelism with the computation of the previous
section is more transparent if one assumes, as we do,
that $f_{ij}$ is skewsymmetric. In~\cite{moros} a 
particular instance of this situation was considered, 
but they assumed that the variables $\dot x$ are 
functions of the $x^i$'s, a hypothesis without much 
physical justification, they assume a special form
of the $f_{ij}$'s which is completely unnecessary, 
and they place their argument in the constant 
noncommutative case.

In other words, we are now  replacing~\eqref{eq:Ptensor} 
by the general Poisson tensor 
$$
\Lambda= g_{ij}(x,\dot x) \, \pd{}{x^i}\w \pd{}{x^j} 
+\frac{1}{m}\bigl(\dl_{ij} + f_{ij}(x,\dot x)\bigr)
\,\pd{}{x^i}\w \pd{}{\dot x^j}
+ A_{ij}(x,\dot x)\,\pd{}{\dot x^i}\w \pd{}{\dot x^j}\,, 
$$  
and the problem is to determine the functions $A_{ij}$
and the Hamiltonian $H$ given the functions $g_{ij}$ and 
$f_{ij}$, and the equations of motion 
\eqref{eq:fmotioneq} and \eqref{eq:smotioneq}, or 
Newton's equation
$$
m\ddot x^j = F^j(x,\dot x).
$$

Once more, when applying the vector field $\Gamma$ 
to~\eqref{eq:GPbracket} we obtain
$$
\Gamma {f_{ij}}
= m\{\dot x^i,\dot x^j\} +\{x^i,F^j\},
$$
therefore 
$$
\{\dot x^i,\dot x^j\}
=\frac{1}{m}(\Gamma{f_{ij}}-\{x^i,F^j\})\,,
$$ 
and since $f_{ij}$ is skewsymmetric, 
$$
\{x^i,F^j\}= -\{x^j,F^i\},
$$
so a field $B$ can be defined as in~\eqref{eq:defB}
or~\eqref{eq:defBbis}, and exactly the same computations 
can be performed, leading to some equations a bit more 
involved, but similar to~\eqref{eq:fMeqbis} 
and~\eqref{eq:sMeqbis}. We see no point
in repeating the calculations. 

In this context the equations of motion become
\begin{align*}
\dot x^i &=\{x^i,x^j\}\pd{H}{x^j}
+ \{x^i,\dot x^j\} \pd{H}{\dot x^j} \, , \\
F^i  &=\{\dot x^i,x^j\}\pd{H}{x^j}
+ \{\dot x^i,\dot x^j\} \pd{H}{\dot x^j}\, ,
\end{align*}
which are more complicated than the classical
ones, but, in principle, a Hamiltonian 
description is still possible in the 
noncommutative setting.

We conclude that noncommutativity does allow 
other dynamics than the standard formalism of 
electrodynamics.

\subsection*{Acknowledgments}

We thank Jos\'e ~Gracia-Bond\'{\i}a and 
Giuseppe Marmo for useful conversations. JFC  acknowledges financial support    from research projects BFM2003-02532 and DGA-GRUPOS CONSOLIDADOS 225-206.
HF thanks the Departamento de F\'{\i}sica Te\'orica de la 
Universidad de Zaragoza for its hospitality and
acknowledges support from the Vicerector\'{\i}a de 
Investigaci\'on of the Universidad de~Costa~Rica.

\end{document}